\documentclass[journal=jpclcd,manuscript=article]{achemso}

\usepackage{chemformula} 
\usepackage[T1]{fontenc} 
\usepackage{bm}

\author{Nicolas Gastellu}
\affiliation{Department of Chemistry, McGill University, Montr\'{e}al, Qu\'{e}bec, Canada}
\author{Ata Madanchi}
\affiliation{Department of Physics, McGill University, Montr\'{e}al, Qu\'{e}bec, Canada}
\author{Lena Simine}
\email{lena.simine@mcgill.ca}
\affiliation{Department of Chemistry, McGill University, Montr\'{e}al, Qu\'{e}bec, Canada}
\alsoaffiliation{Department of Physics, McGill University, Montr\'{e}al, Qu\'{e}bec, Canada}

\title{Disentangling morphology and conductance in amorphous graphene}

\begin{document}

\begin{tocentry}
\begin{center}
\includegraphics[width=1.0\textwidth]{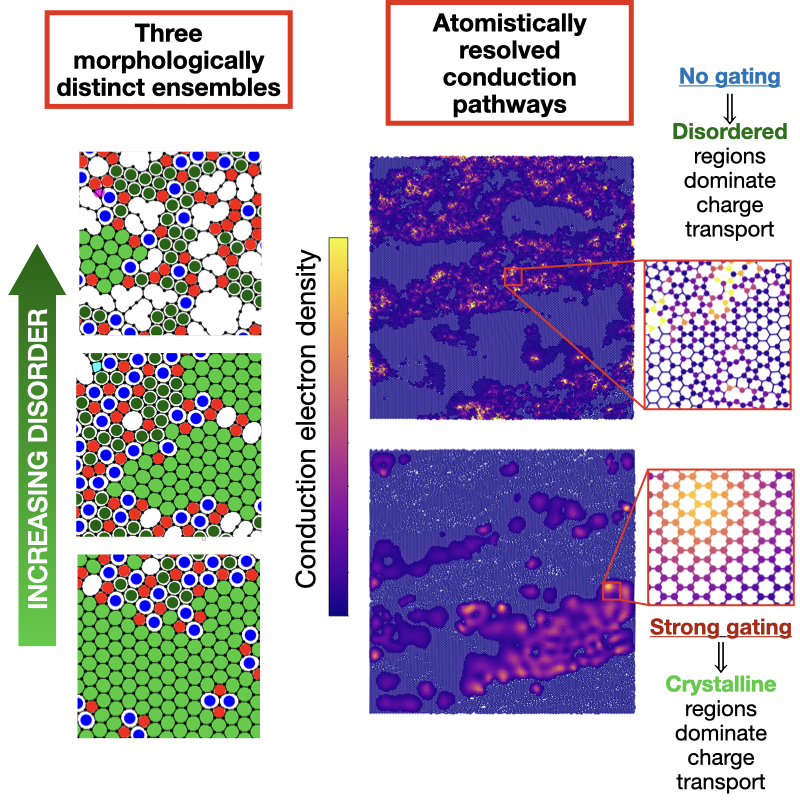}
\end{center}
\end{tocentry}


\begin{abstract} 
	Amorphous graphene or amorphous monolayer carbon (AMC) is a family of carbon films that exhibit a surprising sensitivity of electronic conductance to morphology. We combine deep learning-enhanced simulation techniques with percolation theory to analyze three morphologically distinct mesoscale AMCs. Our approach avoids the pitfalls of applying periodic boundary conditions to these fundamentally aperiodic systems or equating crystalline inclusions with conducting sites. We reproduce the previously reported dependence of charge conductance on morphology and explore the limitations of partial morphology descriptors in witnessing conductance properties. Finally, we perform crystallinity analysis of conductance networks along the electronic energy spectrum and show that they metamorphose from being localized on crystallites at band edges to localized on defects around the Fermi energy opening the possibility of control through gate voltage. 
 \end{abstract}



Understanding the role of microscopic structure in determining the macroscopic properties of a material is an important overarching goal in chemistry, physics, and molecular engineering. Of particular interest is establishing structure-function relationships for materials classified as ``amorphous''. Although the adjective ``amorphous'' may suggest that the structure is simply a random arrangement of atoms, it is well known that the properties of real amorphous materials sensitively depend on the preparation route (deposition rates, temperature of substrate, etc.) implying configurational tunability at the atomic-level. The rich chemical and conformational landscapes of these materials pose many challenges to theory and experiment, and drive innovation in both~\cite{review1}. 

Disorder inherent in amorphous materials is difficult to control and characterize experimentally and it is hard to replicate in a simulation~\cite{review1}. Amorphous graphene, or amorphous monolayer carbon (AMC) has recently emerged as a uniquely suitable model system for studying structure-function relationships in amorphous materials~\cite{toh2020synthesis}. Thanks to the two-dimensional configuration which can be visualized using microscopy and to the conjugated-carbon composition that is relatively easy to model computationally the correspondence between simulation and experiment can be established precisely. In recent years AMCs have garnered significant research interest regarding the effect of disorder on the electronic~\cite{kapko2010electronic,van2012insulating,lherbier2013highly}, thermal~\cite{zhu2016phonons,antidormi2020thermal}, structural~\cite{kumar2012amorphous,zhang2022structure,chen2021stone}, and mechanical properties~\cite{felix2020mechanical}.
However, it is the novel synthetic protocol by Tian \textit{et al.} that turned AMCs into a class of amorphous materials resolved along a well characterized morphological axis~\cite{tian2023disorder}. 

\begin{figure}
  \centering
  \includegraphics[width=\textwidth]{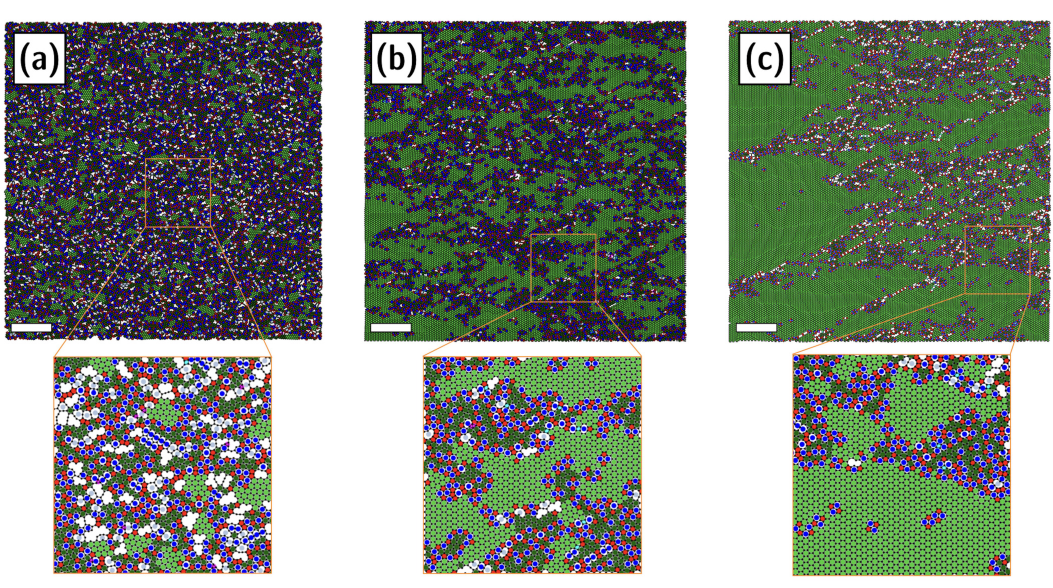}
  \caption{\textbf{MAP-generated AMCs}.  To highlight the topological composition of AMC samples, pentagons are highlighted in red, isolated hexagons are in dark green, crystalline hexagons are in lime green (see Section S2 of Supporting Information for a technical definition of isolated and crystalline hexagons), 7- and 8-membered rings are in dark blue, 9- and 10-membered rings are in grey, squares are in turquoise, and triangles are in pink.
  \textbf{(a)} Example structure from the sAMC-500 ensemble.
  \textbf{(b)} Example structure from the sAMC-q400 ensemble. 
  \textbf{(c)} Example structure from the sAMC-300 ensemble. 
  \textit{Scale bars}: 5~nm for the main figures, and 1.5~nm for the insets, which correspond to magnified regions of the full structures.}
  \label{fig:ensembles}
\end{figure}

Our focus here is on electrical conductance in AMCs, a property which was shown to have remarkable sensitivity to morphology. It was reported that a small change in synthetic conditions of AMC films led to a dramatic billion-fold increase in the sheet conductance~\cite{tian2023disorder}. Building a mathematical relationship between morphology and conductance is a way to gain insight into this instability. From the computational perspective, however, modeling charge transport in such systems is extremely challenging because precise atomistic AMC structures need to be analyzed at the mesoscale, the scale of experimental measurements. Since AMC is a covalently-bonded network of conjugated carbons and cannot be meaningfully fragmented, a good model must generate and analyze high quality atomistically resolved mesoscale samples. By anchoring computational results in experimental data, this paper aims to unravel the morphology-conductance relationship in AMCs by using machine-learning-augmented simulation methodologies~\cite{kilgour2020generating,madanchi2024simulations,deringer2017GAP}, thus demonstrating that computational modeling can be a helpful tool in amorphous materials research.  




For the purposes of our analysis we generate three ensembles of morphologically-distinct mesoscale AMCs: two to be compared directly with experimentally reported data, and one to be used to demonstrate an ambiguity in the currently accepted AMC morphology classification method. Samples of the three morphological classes are shown in Figure~\ref{fig:ensembles}. Each mesoscopic AMC has the dimensions of roughly 40~nm~$\times$~40~nm and was generated using a sampling approach based on generative machine learning: the Morphological Autoregressive Protocol (MAP)~\cite{kilgour2020generating,madanchi2024simulations}. 
This approach takes advantage of decaying structural correlations characteristic of amorphous materials to sample large-scale conformations by extrapolating from smaller (order of correlation length) samples. The training samples in this work were produced by a Monte-Carlo bond switching protocol driven by ML energies from the C-GAP-17 potential~\cite{deringer2017GAP}. In order reduce the error bars on computational predictions to acceptable levels more than 200 samples were generated in each ensemble. 
For full technical details on dataset generation and the MAP implementation, see Section S1 of Supporting Information.

\begin{figure}
  \centering
  \includegraphics[width=0.6\textwidth]{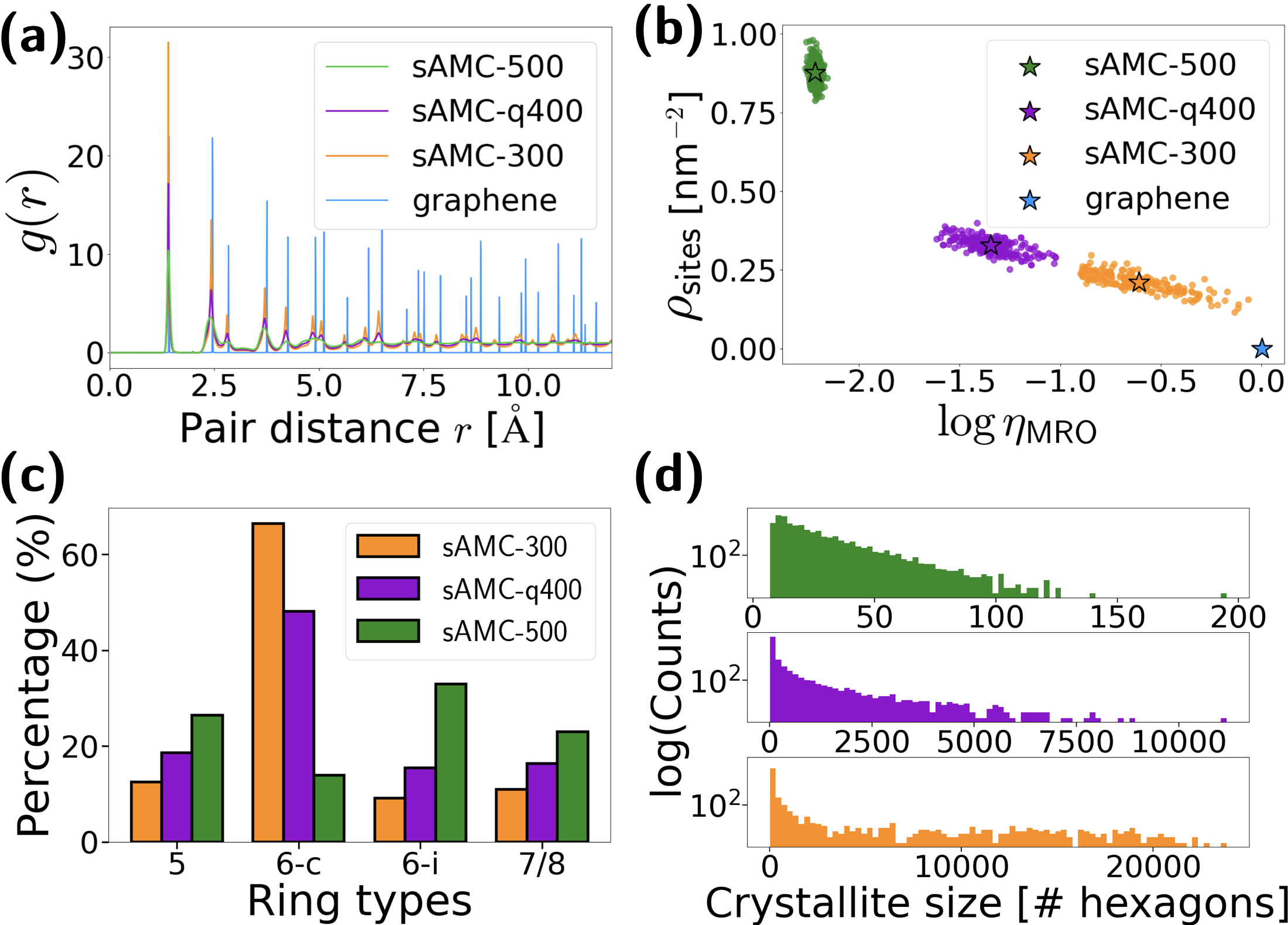}
  \caption{\textbf{Characterisation of disorder in AMC}. \textbf{(a)} Ensemble-averaged pair correlation functions of sAMC-500 (green), sAMC-q400 (purple), and sAMC-300 (orange). The pair correlation function $g(r)$ of pristine graphene (blue) is also shown as a reference.
  \textbf{(b)} Scatter plot of all the structures in the three ensembles in $(\log\eta_{\text{MRO}}, \rho_{\text{sites}})$ space. The green, purple, and orange stars mark the respective positions of the sAMC-500, sAMC-q400, and sAMC-300 ensemble averages. The position of a 40~nm$\times$40~nm graphene flake is shown as a blue star for reference. In this space sAMC-300 is closer to graphene than to the highly disordered sAMC-500.
  \textbf{(c)} Carbon ring size distributions in generated AMC ensembles. Crystalline and isolated hexagons are respectively denoted 6-c and 6-i.
  \textbf{(d)} Distribution of crystallite (i.e. cluster of crystalline hexagons) sizes in sAMC-500 (top, green), sAMC-q400 (middle, purple), and sAMC-300 (bottom, orange) structures, plotted on a logarithmic scale.}
  \label{fig:structural_characterisation}
\end{figure}

To classify the morphology we use the same descriptors as Tian \textit{et al.}: (i) the pair correlation function $g(r)$ averaged over each ensemble (see Figure \ref{fig:structural_characterisation}a); (ii) the short- and medium-range order parameters, $\rho_{\text{sites}}$ and $\log\eta_{\text{MRO}}$ defined below, (iii) the distribution of different types of carbon rings, and (iv) the area distribution of crystalline inclusions. The order parameter $\rho_{\text{sites}}$ is the number of connected clusters of undistorted carbon atoms per unit area in a given structure.
A carbon atom is considered undistorted if the bond lengths and bond angles it forms with its immediate neighbours deviate by under 10\% from those in pristine graphene.
This order parameter therefore measures the persistence of short-range order in a AMC sample.
The medium-range order parameter $\log\eta_{\text{MRO}}$ order parameter is related to the integral of the pair correlation function in the medium range distance between 4~\AA~ and 12~\AA. For technical details on all four descriptors, see Section S2 of Supporting Information.
A discussion of our ensembles' bond length and bond angle distribution can also be found in Section S3 therein.

We start our morphology analysis with the most disordered ensemble, see Figure~\ref{fig:ensembles}a for sample visualization, and the green line in Figure~\ref{fig:structural_characterisation}a for evidence of fast-decaying pair correlations (note that unlike in Ref~\citenum{tian2023disorder}, we normalise our pair correlation functions such that $g(r)\xrightarrow[r\rightarrow\infty]{}1$). As shown in Figure~\ref{fig:structural_characterisation}b, the green set of points corresponding to this ensemble, it is positioned in the upper left corner of the $\log\eta_{\text{MRO}}$  vs. $\rho_{\text{sites}}$ plot -- close to the position of AMC-500 ensemble reported by Tian \textit{et al.}~\cite{tian2023disorder}, and its distribution of rings shown Figure~\ref{fig:structural_characterisation}c (in green) is close to the experimental AMC-500 ensemble as well. 
Due to its similarity in morphological metrics to the AMC-500 from Ref.~\citenum{tian2023disorder}, we will refer to it as ``simulated AMC-500'', or sAMC-500 henceforth.

The next ensemble we discuss is the most grahene-like ensemble shown in Figure~\ref{fig:ensembles}c. It is much more ordered as it contains large sections of almost perfect hexagonal order (graphene-like hexagonal motifs are highlighted in lime green in Figures~\ref{fig:ensembles}a-c). 
This ensemble displays ring statistics that are close to the AMC-300 samples in Tian \textit{et al.}~\cite{tian2023disorder} and  it is found in a similar region of the $\log\eta_{\text{MRO}}$  vs. $\rho_{\text{sites}}$ space. We will therefore refer to this ensemble as ``simulated AMC-300'', or sAMC-300.

The third ensemble, visualized in Figure~\ref{fig:ensembles}b, is somewhat difficult to classify. Similarly to Tian \textit{et al.}'s AMC-400, our third set of structures exhibits a degree-of-disorder greater than that of the sAMC-300 samples, but lower than that of those in the sAMC-500 as inferred from observations of disorder in Figures~\ref{fig:ensembles}a-c, and the intermediate decay of correlations in Figure~\ref{fig:structural_characterisation}a. It is found in a position similar to AMC-400 in $(\log\eta_{\text{MRO}},\rho_{\text{sites}})$ space.
However, the ring statistics (Figure \ref{fig:structural_characterisation}c in purple) show a balance of crystalline hexagons (6-c) vs. non-crystalline hexagons (6-i) that is reversed with respect to the experimentally generated AMC-400 samples (see Figure~2h in Ref.~\citenum{tian2023disorder}): in our simulated structures, the crystalline hexagons 6-c are much more common than non-crystalline 6-i hexagons, while the opposite is true in the experimental samples. 
Because of this mismatch, we will refer to this simulated ensemble as sAMC-q400 (for ``quasi-400''). This ensemble highlights the possibility of non-unique classification of morphologies based on position in the ($\log\eta_{\text{MRO}}$, $\rho_{\text{sites}}$) space. 


It is widely accepted that the lack of periodicity in amorphous materials leads to spatially localized electronic states~\cite{anderson1958absence,thouless1974electrons,lee1985disordered}.
This phenomenon prevents charge carriers from travelling coherently across the material.
Instead, charge transport in amorphous phases is modelled within a variable range hopping (VRH) picture, wherein charge carriers incoherently travel between localised sites by exchanging energy with a bath of molecular vibrations~\cite{mott1971electronic}.
While the billionfold enhancement in conductance displayed by the high-crystallinity AMC samples in reference~\citenum{tian2023disorder} might lead one to suspect that a coherent band transport mechanism may dominate, transport measurements show that their resistance-temperature relation obeys the two-dimensional Mott law
\cite{tian2023disorder,mott1971electronic,toh2020synthesis}, which is typical of VRH in systems with a slowly varying density of states (DOS)~\cite{mott1971electronic,ambegaokar1971hopping,shklovskii1984electronic}.
This is strong evidence that the VRH picture remains valid, even in for such ordered AMCs.  
The hopping sites in the VRH picture are usually taken to be spatially localised eigenstates of the electronic hamiltonian~\cite{shklovskii1984electronic}.
Since AMCs are $sp^2$-hybridized conjugated carbon systems~\cite{toh2020synthesis}, it may be safely assumed that only the $\pi$-network electrons contribute to conduction. We therefore model the electronic structure of mesoscopic AMC fragments using an all-atom tight-binding hamiltonian $\mathcal{H} = \sum_{\langle i,j\rangle} t_{ij} (|\varphi_i\rangle\langle\varphi_j| + |\varphi_j\rangle\langle\varphi_i|)$, where $|\varphi_i\rangle$ denotes the $2p_z$ orbital centered on the $i$th carbon in the AMC sample. The sum is carried out over all nearest-neighbour pairs $\langle i,j\rangle$, and the semi-empirical parametrisation of the hopping elements $t_{ij}$ is adapted from prior works~\cite{sterpone2008molecular,jailaubekov2013hot}, see Section S4 of Supporting Information for details.

We work with 40~nm$\,\times\,$40~nm atomistically resolved samples, and thus do not resort to coarse-graining, applying periodic boundary conditions, modeling only partially resolved structures, or any other kind of commonly employed simplifications. 
While the enormous size of the hamiltonian matrices and the large number of samples in our ensembles preclude us from resolving the full eigenspectrum using standard numerical routines, doing so is not necessary for this problem. 
Since thermally activated conductance is expected to be mediated by a relatively narrow band of thermally accessible states, we partially diagonalise each hamiltonian using the Lanczos algorithm~\cite{golub1996matrix}, which is well-suited to the tight-binding hamiltonians' sparse structure (Section S5 of Supporting Information). We thereby obtain the subset of molecular orbitals (MOs) whose energies lie within $4k_{\text{B}}T$ above the chemical potential $\mu$ at room temperature ($T = 300$~K), under different gating conditions.
We focus on three regimes: (i) the ungated regime in which $\mu = \epsilon_F$, the sample's Fermi level at half-filling; (ii) the regime where a strong negative gate voltage is applied and $\mu = \epsilon_0$, the smallest eigenvalue of $\mathcal{H}$; and (iii) the regime where a strong positive gate voltage is applied and $\mu = \epsilon_N - 4k_{\text{B}} T\,$ where $\epsilon_N$ is the greatest eigenvalue of $\mathcal{H}$.

It is expected that solutions obtained from such a low level of theory may possess some degree of artefacts. In this case, we noticed that some of the eigenstates we obtain are delocalised in ways that we do not expect to withstand the effects of decoherence and localization due to factors that were not included in our calculation but are certainly present in the real system. Some of the factors that are not included in our model are the electron-electron and electron-phonon interactions, as well as the interactions with substrate, and distortions such as ripples or buckling that AMC sheets are expected to undergo at ambient conditions~\cite{meyer2007structure,ma2009stone}, which would disrupt the extent of $sp^2$ conjugation and contribute to electronic localisation~\cite{krasavin2022electrical}. In our case, the delocalized eigenstates/molecular orbitals (MOs) which we regarded as pathological typically have several disjoint pockets of high electronic density separated by distances that can span 10's of nanometres -- almost the entire AMC structure's length (see Figs.~S3 and S4 in Supporting Information). This behavior was more common among MOs found near $\epsilon_F$, whereas the eigenstates at the band edges expectedly tended to be localised, with very few exceptions. In order to model VRH of charges we have made the choice to construct conduction space in which we artificially localize the pathological MOs into multiple disjoint charge-hopping sites.

To construct the state space for the VRH network, we have developed a procedure based on $k$-means clustering to extract hopping sites from the MOs supported by each AMC fragment, (see section S7 of Supporting Information). 
This approach partitions each (n$^{th}$) MO $|\psi_n\rangle$ into a set of $m_n$ localised states $\{|s_{n,i}\rangle\}_{i=1,\ldots, m_n}$, from which site positions $\{\mathbf{R}_{n,i}\}_{i=1,\ldots,m_n}$ and delocalisation radii $\{a_{n,i}\}_{i=1\ldots, m_n}$ can be extracted.
MOs that are already localised are left unchanged by this procedure.
As the last correction mechanism for numerical artifacts in the electronic structure modeling, the occasional sites whose effective area $\pi a_{n,i}^2$ exceeded the area of the largest crystalline inclusion found in the ensemble (Fig.~\ref{fig:structural_characterisation}d) were considered numerically artificial and removed from our calculation. 
We tabulate the maximum allowed site radii $a_{max}$ for the three AMC ensembles in Table~\ref{tab:results}.
We note that the experimental crystallite sizes reported in reference~\citenum{tian2023disorder} are much smaller than the ones we obtained from our simulated structures. This may be because the AMC microscopy images processed by Tian \textit{et al} are much smaller ($\sim$~5~nm~$\times$~5~nm) than the sAMC samples we use in our analysis ($\sim$~40~nm~$\times$~40~nm). 

Having constructed the VRH space, we apply percolation theory to estimate AMC conductances.
Percolation theory has been very successfully applied as an analytical framework to estimate the VRH conductance of various classes of disordered semiconductors whose density of states (DOS) has a simple closed-form expression (e.g. a Gaussian distribution)~\cite{ambegaokar1971hopping,ovskit1971impurity,pike1974percolation,grunewald1979hopping,vissenberg1998theory,baranovskii2002percolation,rubel2004concentration,zvyagin2008percolation,rodin2011hopping,nenashev2013advanced}. 
Practically, percolation-based approaches are attractive because they do not suffer from stability or convergence issues associated with the usual techniques of VRH simulations like the solution of a transport master equation~\cite{yu2001molecular,pasveer2005scaling}, or Monte Carlo sampling of hopping trajectories~\cite{bassler1993charge,willson2023jumping}. 
Seeing as the DOS profiles for the three morphological classes we consider are not known analytically, we developed a numerical implementation of percolation of theory, which yields an ensemble-averaged estimate for charge conductance.

We define the hopping rate $\omega_{ij}$ between sites $i$ and $j$ using a Miller-Abrahams expression which we modified to accommodate for variable site radii: $\omega_{ij} \sim \,e^{-\xi_{ij}}$, where the dimensionless quantity $\xi_{ij}$ can be thought of as an effective distance between sites $i$ and $j$ and depends on the sites' positions, energies, and radii (see equation (S5) for $\omega_{ij}$ in Supporting Information).
Each structure admits a critical distance $\xi_c$ at which a cluster of sites obeying $\xi_{ij}\le\xi_c$ percolates the sample by connecting its right edge to its left edge.
The randomness inherent to the AMC fragments, as well as their finite size, will lead $\xi_c$ to fluctuate from fragment to fragment.
Following an approach similar to Rodin and Fogler~\cite{rodin2011hopping}, we estimate each ensemble's conductance $\sigma$ as follows:

\begin{equation*}
  G = \frac{q_e^2\omega_0}{k_{\text{B}} T}\int\text{d}\xi e^{-\xi} P(\xi)\,,
\end{equation*}

where $P(\xi)$ is the probability of having a percolating cluster through sites obeying $\xi_{ij} \le \xi$ in a given ensemble, $\omega_0=1$~fs$^{-1}$ is the escape frequency, $q_e$ is the elementary charge, $k_{\text{B}}$ is Boltzmann's constant, and $T = 300$~K denotes temperature. See Section S8 of Supporting Information for technical details. 

Next, we summarize and discuss the results of our modeling. First, we reproduce experimentally observed conductances reported in Ref.~\citenum{tian2023disorder} for sAMC-500 and sAMC-300 ensembles. 
We then discuss the deviation of the predicted conductance for the sAMC-q400 ensemble relative to the experimentally characterized in Ref.~\citenum{tian2023disorder}  AMC-400. 
Our results imply that a unique map between the $(\log\eta_{\text{MRO}},\rho_{\text{sites}})$ space and conductance, in contradiction to previous assertions (e.g., Figure 4a in Ref.~\citenum{tian2023disorder}), does not exist. 
We conclude with the discussion of the possibility of controlling conductance in AMCs by applying gate voltage and in effect modifying the characteristics of the charge transport pathways while keeping the AMC sample unchanged.

\begin{figure}
  \centering
  \includegraphics[width=0.6\textwidth]{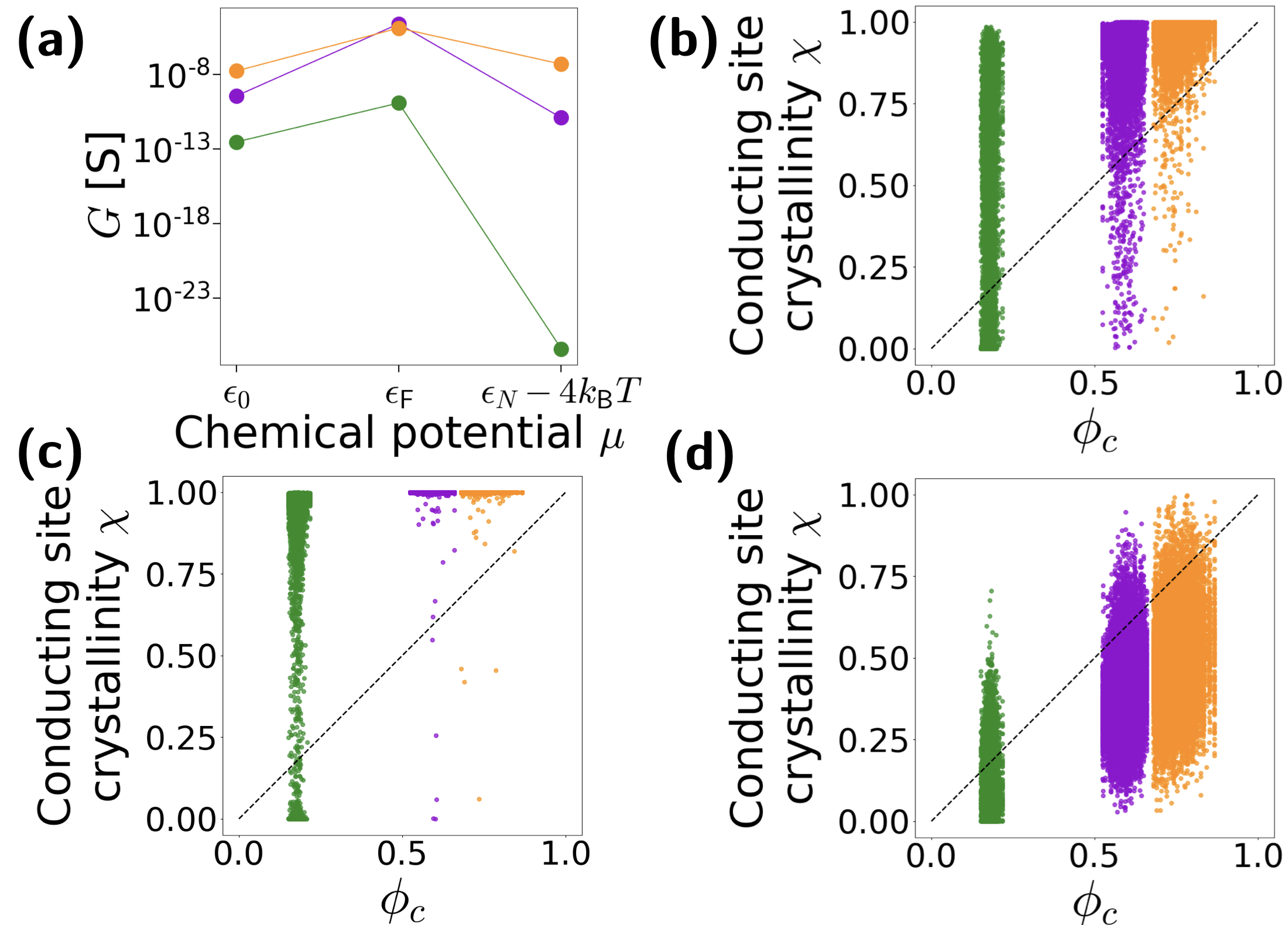}
  \caption{\textbf{Electronic conductance and the morphology of conducting pathways in AMC}. \textbf{(a)} Electrical conductance $G$ of the three sAMC ensembles, at different gate voltages. The error bars are omitted because they are too small to be visible (see Table~\ref{tab:results}).
  \textbf{(b)}-\textbf{(d)} Crystallinity $\chi$ of the electron hopping sites in a conduction network, plotted against the fraction of crystalline atoms $\phi_c$ in the AMC samples under the three gating regimes: \textbf{(b)} $\mu = \epsilon_0$, the smallest eigenvalue of electronic Hamiltonian, \textbf{(c)} $\mu = \epsilon_{N} - 4 k_{\text{B}}T$, where $\epsilon_N$ is the largest eigenvalue of electronic Hamiltonian, and \textbf{(d)} $\mu = \epsilon_F$. Sites from sAMC-500 samples are shown in green, sites from sAMC-q400 samples are in purple, and sites from sAMC-300 samples are in orange. Sites that are clustered high above the diagonal are localized on crystalline regions while those those clustered under the diagonal show preference for defects.}
  \label{fig:results}
\end{figure}


We performed our conductance calculations at $T = 300\,$K, under different gating conditions. The results are plotted in Figure~\ref{fig:results}a, and a summary is tabulated in Table~\ref{tab:results}.
First, we focus on the conductance in the ungated case in which charge transport is carried by states close to the Fermi energy ($\epsilon_F$) of the AMC, the middle set of data-points in Figure~\ref{fig:results}a. The two ensembles that we may directly compare to experimental data are sAMC-300 shown in orange and sAMC-500 shown in green. We observe a conductance gap of $5$ orders of magnitude between the two in our calculations which matches well the gap observed in Ref.\citenum{tian2023disorder}, see Figure~3d therein\footnote{Note that all data plotted Figure~3d of Ref.~\citenum{tian2023disorder} were collected at room temperature~\cite{email_liu}. Its abscissa thus corresponds to the substrate temperature during the AMC films' growth, which controls their degree of disorder, rather than the temperature of the films during the resistivity measurements.}.

\begin{table}[h]
  \centering
  \begin{tabular}{lccc}
    \hline
                                        & sAMC-500 & sAMC-q400  & sAMC-300 \\
    \hline
  $a_{max}$ {[}\AA{]}                   & 18.03 & 136.47 & 199.33 \\
  $p_{6c}$ {[}\%{]}                     & 13.64 & 48.11  & 66.41    \\ 
  $G(\mu = \epsilon_{0})$ {[}S{]} & $(2.99\pm 2.43)\cdot 10^{-13}$  & $(3.64\pm 1.60)\cdot 10^{-10}$ & $(1.72\pm 1.20)\cdot 10^{-8}$   \\
  $G(\mu = \epsilon_F)$ {[}S{]}  & $(1.24\pm 0.11)\cdot 10^{-10}$  & $(2.51\pm 0.40)\cdot 10^{-5}$ & $(1.30\pm 0.15)\cdot 10^{-5}$  \\
  $G(\mu = \epsilon_{N} - 4k_{\text{B}}T)$ {[}S{]}   & $(4.11\pm4.03)\cdot 10^{-27} $  & $(1.27\pm 1.26)\cdot 10^{-11}$ & $(5.00\pm 3.32)\cdot 10^{-8}$   \\
  \end{tabular}
  \caption{\textbf{Summary of transport properties and structural characteristics exhibited by the simulated AMC ensembles.} For each ensemble, $a_{max}$ is the effective radius of the largest crystalline inclusion (Fig. ~\ref{fig:structural_characterisation}~d) and the delocalisation radius cutoff used in the percolation calculation; $p_{6-c}$ is the percentage of crystalline hexagons (based on Fig. ~\ref{fig:structural_characterisation}~c); and $G(\mu=E)$ denotes conductance when gating sets each structure's chemical potential $\mu$ to energy $E$.}
  \label{tab:results}
  \end{table}

Of particular interest is the high conductance of the sAMC-q400 ensemble which is found to be on par with the high, relative to other AMCs, conductance of sAMC-300. The sAMC-q400 ensemble falls close to the experimental AMC-400 ensemble in the $(\log\eta_{\text{MRO}},\rho_{\text{sites}})$ space  with the center close to (-1.3,0.3) point compared to approximately (-1.25,0.35) in the experimental case. However, AMC-400 was experimentally and theoretically shown to be a perfect insulator. Previous modeling approaches assign a smooth map from the $(\log\eta_{\text{MRO}},\rho_{\text{sites}})$ space to conductance with vanishing values in this particular region of the $(\log\eta_{\text{MRO}},\rho_{\text{sites}})$ space, and would therefore struggle to reconcile the sharp difference between the behaviors of AMC-400 and sAMC-q400. Our result, on the other hand, demonstrates that the morphology-conductance relationship in AMC is more complex than previously thought. The morphological difference between sAMC-q400 and AMC-400 is visible in ring-distribution statistics (Figure~\ref{fig:structural_characterisation}c).



Finally, we discuss the possibility to weakly decouple the conductance from morphology in AMCs. The idea behind this is that molecular orbitals in different regions of the spectrum tend to have differing morphological characteristics. We previously discussed the emergence of edge-states around the middle of the spectrum and bulk-localization of MOs towards the edges of the ``band'' in amorphous graphene nanoflakes~\cite{gastellu2022electronic}. Similar behavior is observed in mesoscale samples of AMCs as well, and by applying gate voltage we, in effect, modify morphological characteristics of the conducting VRH networks while keeping the overall atomistic morphology fixed.

We note that the effects of applying a gate voltage in our simplified modeling reflect only the characteristics of the MOs that carry the current under the different conditions - the changes to resistance of contacts for example are neglected. Under this idealized assumption, we observe that conductance depends rather strongly on gating and that more disordered AMCs show larger sensitivity. To understand the origin of this effect we need to quantify the morphological changes, i.e., the structural metamorphosis, that conducting networks undergo when a gate voltage is applied. 


The connection between the morphology and conductance can be clarified by focusing on the structure of only those regions of AMC samples on which the electronic states that dominate charge transport are localized.
To do so we define the crystallinity $\chi$ of a given VRH state $|\psi\rangle$ as follows:
\begin{equation}
\chi(|\psi\rangle) = \sum_{j\in\mathcal{C}} |\langle\varphi_j|\psi\rangle|^2\,,
\end{equation}
where $\mathcal{C}$ corresponds to the set of crystalline atoms -- i.e. atoms belonging to a crystalline hexagon (previously referred to as 6-c).
In words, $\chi(|\psi\rangle)$ corresponds to the aggregated density of $|\psi\rangle$ which lies on the crystalline regions of a given AMC sample.

Figures~\ref{fig:results}b-d show the crystallinity of each VRH site belonging to a percolating cluster against the fraction of crystalline atoms $\phi_c$ in the relevant AMC sample.
Focusing first on the gated regimes depicted in Figures~\ref{fig:results}b-c, we find that conducting sites at the band edges (i.e. far from $\epsilon_F$), tend to be localized preferentially on crystalline inclusions since they tend to lie above the diagonal and thereby satisfy $\chi>\phi_c$, meaning that they are more crystalline than the rest of the structure. We infer from this that charge transport in these regimes is mostly carried by hopping from one crystalline site to another, giving a partial rationalization to an intuitive and common modeling assumption.
 

The picture changes when we move to the ungated case shown in Figure~\ref{fig:results}d.  There, we find that unlike the states at the band's edges, the mid-spectrum eigenstates produce VRH sites which predominantly satisfy $\chi < \phi_c$ (Fig. ~\ref{fig:results}d), i.e., that are preferentially found on defects in all three ensembles. 
Charge hopping in the ungated regime therefore takes place predominantly over the disordered regions of AMC. Within the framework of VRH, the bias of the conduction networks towards disordered regions puts the propensity of defects in competition with the localization of the electronic states. The interplay between these two properties gives rise to the predicted conduction trends, and originates in the dependence of the inter-site hopping rates $\omega_{ij}$ on the sites' radii and site-site distances (equations (S5) and (S6) in Supporting Information). For instance, our calculations show that sAMC-q400 exhibits ungated conductance close to sAMC-300 in spite of being much more disordered overall. This happens because it strikes a balance between the two competing effects: it contains more extensive disordered regions to increase the density of hopping sites relative to sAMC-300, while also retaining enough structural order to give rise to sites with reasonably large radii.

Figure~3 also highlights the contrast between the crystallinity distributions of conducting sites in the different sAMC ensembles, especially in the strong gating regimes (Figs.~3b-c). 
In this respect, the ordered ensembles sAMC-300 (orange points in Figs.~3b-c) and sAMC-q400 (purple) are qualitatively very similar: their low- and high-energy conducting sites are clustered near the top of Figures 3b-c -- indicating their highly crystalline character -- while the extremal energy sites in sAMC-500 (green) follow a much more uniform crystallinity distribution.
We discuss such qualitative differences in the crystallinity distributions of our three ensembles' conducting sites in greater detail in Section S9 of the Supporting Information, but comment on them briefly here.
First, we note that the crystallinity of a structure's conducting sites is strongly influenced by the crystallinity of the energy eigenstates from which they originate (see Figure~S5 in the Supporting Information).
Starting with sAMC-500, its structures are akin to continuous random networks (CRNs), which are statistically homogeneous~\cite{zachariasen1932atomic}.
The MOs supported by sAMC-500 structures will therefore tend to evenly sample a diverse set local atomic environments~\cite{dong1998atomistic,ludlam2005universal}.
These MOs therefore exhibit a correspondingly diverse set of crystallinity values (cf. Fig.~S5 in the Supporting Information), which in turn leads to the more uniform nature of the crystallinity distributions of their corresponding hopping sites (Figs~3b-c).
Conversely, sAMC-300 and sAMC-q400 structures feature extended crystalline domains, and universally obey $\phi_c > 0.5$.
These two ensembles are therefore much more similar to defected graphene than to a CRN.
Defected semiconductors are known to exhibit defect-localised states near $\epsilon_F$, while the states deep within the occupied and virtual manifolds will retain a highly crystalline character~\cite{shklovskii1984electronic,mott2012electronic}.
This is entirely consistent with what we observe in Figure S5, and therefore explain the very high crystallinity of their hopping sites.



Under our simplifying assumptions, the gate-voltage modulation charge transport in AMCs is owed to the metamorphosis of the conducting states along the electronic energy spectrum.
This ready and reversible tunability of AMC's conductance -- taken together with the low heat conductance which is owed to its inherently disordered bonding network~\cite{antidormi2020thermal} -- make is potentially  an attractive candidate for thermoelectric applications. This observation is likely to remain at the conceptual level until effective strategies are invented to increase conductivities in AMCs to levels sought after in thermoelectric materials, i.e. $\sim 10$~S~cm$^{-1}$~\cite{liu2020n}. In the past, nitrogen doping has been shown enhance the electrical conductance of AMC by an order of magnitude~\cite{zhao2012synthesis} and sets a precedent for future developments in this space.


In this work we have combined deep learning-enhanced simulation techniques with percolation theory to model charge conductances in three morphologically distinct mesoscale AMCs. We have overcome the challenges of modeling electronic conductance in mesoscale atomistically resolved covalently bonded networks of conjugated carbons by developing a custom partial diagonalisation procedure based on the Lanczos algorithm and adapted the percolation theory calculation of charge conductance to the pecularities of the AMC system. Our protocol is noteworthy in that is avoids the artifacts that may arise from applying periodic boundary conditions to aperiodic amorphous structures (a common practice), or from oversimplifying assumptions regarding transport mechanism. We reproduce the reported dependence of charge conductance on morphology and discuss the ambiguous relationship between incomplete/partial measures of morphology and conductance. By conducting a crystallinity analysis of the conducting sites we show that they metamorphose in response to gate voltage from being localized on crystallites at band edges to being localized on defects around the Fermi energy.  Inspired by this observation, we explore the possibility to decouple the AMC morphology from electronic conductance by applying a gate voltage and comment on the potential of thermoelectric tunability in amorphous conductors.
\begin{suppinfo}
The following files are available free of charge.
\begin{itemize}
  \item Supporting Information: Detailed description of our methods for (i) modeling AMCs using MAP, (ii) characterising AMC morphologies, (iii) electronic structure calculations, (iv) defining the VRH sites, and (v) computing AMC conductance from numerical percolation theory; Discussion of the bond length and bond angle distributions in our simulated structures; Comparison of our method for diagonalising large tight-binding Hamiltonians with an exact benchmark; Discussion of the crystallinity distributions of tight-binding eigenstates in different of our sAMC structures' energy spectrum. Repositories containing the code used in this work are also included. (PDF).%
\end{itemize}
\end{suppinfo}


\begin{acknowledgement}

We thank Zakariya El-Machachi and Volker Deringer for sharing amorphous graphene dataset they published in Ref.\citenum{deringer2017GAP} and Alessandro Troisi for helpful tips on convergence issues in charge conductance calculations. 
Funding from NSERC Discovery Grant, IVADO, and AI4Design program of the National Research Council of Canada is gratefully acknowledged.   
Computations were performed on Calcul Quebec supercomputers (Narval) funded by Canada Foundation for Innovation (CFI). 
\end{acknowledgement}

\bibliography{refs} 

\clearpage 


\end{document}


\section{S1. Modeling AMCs with the morphological autoregressive protocol (MAP)}

We implement the MAP using the PixelCNN architecture~\cite{van2016conditional,van2016pixel}, a deep generative convolutional neural network. Morphologically distinct ensembles used in our analysis were generated following slightly different protocols. 
For instance, the ensemble sAMC-500 was generated following the protocol reported in reference~\citenum{kilgour2020generating} precisely. 
The training of the MAP model and the generation of ensembles sAMC-300 and sAMC-q400 is detailed below. The relevant code may be found on GitHub~\cite{Gastellu_sAMC_generation}. We note that lack of consistency in generation protocols is immaterial in the context of morphology-function analysis and modeling decisions were based on practical considerations.

The MAP-PixelCNN model was trained on AMC samples ($2\,\text{nm}\times2\,\text{nm}$) which were simulated by a Monte-Carlo bond switching protocol driven by machine-learned energies from the C-GAP-17 potential~\cite{deringer2017GAP} reported in Ref.~\citenum{el2022exploring}. 
Three types of energies were used in the Monte-Carlo acceptance criterion: total energy, atomic energy of the defect pair and atomic energies of the defect pair and the topological nearest neighbours. 
This, in combination with different $\beta$ values allowed sufficient sampling of disordered structures after 10,000 Monte-Carlo steps. In total, 13,944 200-atom fully $sp^{2}$ structures were obtained (just under 2.8 million atomic environments) using this framework with varying levels of disorder. 
For training the MAP ca. 800 structures were selected based on proportion of hexagon content (over $70\%$). 
To prepare the input, the training samples were discretised from real to grid space at a $0.2$~\AA~resolution, empty space was assigned the value 0, carbons were assigned the value 1. 

Overall, the MAP learns the conditional probability distribution for the population of each grid point $i$ given surrounding molecular context: $p(X_i | C_i)$, where the state $X_i$ (carbon or empty space) depends on its local environment $C_i$. This environment, $C_i$, consists of all grid points within a `receptive field' around $i$, capturing the short-to-medium range structural correlations inherent to AMC. To infer the conditional probability the model applies a softmax function to the final layer
\begin{equation}
p(X_i|C_i) =\text{softmax}(z_{i},\tilde{T})
\label{softmax}
\end{equation}
where $z_{i}$ represents the logits predicted for grid point $i$ being empty space or carbon and $\tilde{T}$ is a hyperparameter. The training is done by the stochastic gradient descent to minimize the loss function:
\begin{equation}
    L= -\sum_i\sum_{j= 0,1}{l_{i,j} \log p(X_i=j|C_i)}
\end{equation}
where $l_{i,j}$ is the one-hot encoded value at grid point $i$ of type $j$ (empty or carbon), and $p(X_i|C_i)$ is the prediction made by equation (\ref{softmax}).

By construction, the MAP allows autoregressive generation and extrapolation of molecular structures grid point by grid point, based solely on prior generated molecular fragments.  During the generation process, the softmax function is followed by multinomial sampling using the normalized grid point probabilities to predict the grid point value.  Since in our training samples, the most common motifs are the hexagonal rings, we used the  hyper-parameter $\tilde{T}$ to artificially control the generation of these more abundant motifs - lowering it  reliably produced increasingly ordered (containing larger graphene-like crystalline domains) AMC structures. 
The sAMC-500 ensemble contains 300 samples, the sAMC-q400 ensemble contains 218 samples, and the sAMC-300 ensemble contains 217 samples.  

All structures we lightly post-processed by energy minimization using the AIREBO semiempirical reactive forcefield~\cite{stuart2000reactive}.
The relaxation was carried out using the \textsc{LAMMPS} molecular dynamics software package~\cite{plimpton1995fast, thompson2022lammps}, with leniently defined energy and force tolerance parameters ($10^{-9}$ and $10^{-12}$~eV/\AA, respectively).

\section{S2. Morphological characterization}

\subsection*{Calculating $\rho_{\text{sites}}$}

Following Tian \textit{et al.}~\cite{tian2023disorder}, we define the short-range order parameter $\rho_{\text{sites}}$ as the number of connected clusters of undistorted carbon atoms per unit area, in a given fragment.
As in Ref.~\citenum{tian2023disorder}, a carbon atom is considered undistorted if it is bonded to exactly three neighbours and if the lengths and angles of the bonds connecting it to its neighbours deviate by less then 10\% from the bond length/angle in a pristine graphene lattice ($r_{\text{CC}} = 1.42\,$\AA ~and $\theta = 2\pi/3$).

Using the adjacency matrix representation of each structure (see above section on obtaining the carbon ring distributions of our sAMC structures), and depth-first search traversal of the graph it defines, we identify all clusters composed exclusively of undistorted carbons.
We discard all clusters containing only a single carbon.
We then divide the number $n_c$ of such clusters by the sAMC structure's surface area $A$, which we approximate as a rectangle: $A = L_x\,L_y$, where $L_x = \max_j x_j - \min_j x_j$ (idem for $L_y$).
We thus have: $\rho_{\text{sites}} = n_c / A$.

\subsection{Calculating $\eta_{\text{MRO}}$}

The medium-range order (MRO) parameter $\eta_{\text{MRO}}$ is defined as follows: 

\[ \eta_{\text{MRO}} = A_{\text{MRO}}(\text{AMC ensemble}) / A_{\text{MRO}}(\text{graphene}) \]\,,

where $A_{\text{MRO}}$ is derived from a given structure's pair correlation function $g(r)$: 

\begin{equation*}
  A_{\text{MRO}} = 2\pi\int_{r_1 = 4\text{\AA}}^{r_2 = 12\text{\AA}} |1-g(r)|r\,\text{d}r\,.
\end{equation*}

We compute each sample's pair correlation function by histogramming all interatomic distances smaller than $r_2 = 12$\,\AA.
We use periodic boundary conditions (with the size of the cell chosen to be slightly bigger than the size of the structure at hand along $x$- and $y$-directions) when doing so to avoid artifacts associated with sampling the edges of the structure.
Tian $\textit{et al.}$ used the pair correlation of a physical sample of graphene at room temperature to compute $A_{\text{MRO}}(\text{graphene})$~\cite{tian2023disorder}.
Physical crystals at room temperature tend to have small local distortions (due to phonons), which broadens the otherwise delta function-like peaks in their pair correlation function (this is evident upon inspection of the graphene pair correlation function used by Tian \textit{et al.} in their work -- c.f. Figure 2 l-n of Ref.~\citenum{tian2023disorder}).
These broadened peaks will modify the $A_{\text{MRO}}(\text{graphene})$.

To be able to compare the values of $\eta_{\text{MRO}}$ we obtain from our simulated structures to the $\eta_{\text{MRO}}$ values by Tian \textit{et al.}, we run a $NVT$ molecular dynamics simulation of a graphene flake at $T = 300\,$K, using a timestep of 0.5~fs, and periodic boundary conditions. 
We ran the simulation for 100000 steps and used the structure at step 93000 (i.e. from a part of the simulation that was well-equilibrated) to compute the pair correlation from which we obtained $A_{\text{MRO}}(\text{graphene})$.
We use the \textsc{LAMMPS} package to run the MD simulation.

\subsection*{Obtaining carbon ring distributions}

To find the carbon cycles in a given sAMC structure, we build its graph representation by defining its adjacency matrix as follows:

\begin{align*}
  M_{ij} = \begin{cases}
    1 &\iff \text{ atoms } i \text{ and } j \text{ are covalently bonded,} \\
    0 &\iff \text{ otherwise.}
  \end{cases}
\end{align*}

Here, we consider two atoms as covalently bonded if they are within 1.8~\AA of each other.

We then form a list of all carbon triplets $(i,j,k)$ such that carbons $i$ and $k$ are both bonded to carbon $j$ (but not necessarily to each other).
Each triplet then seeds a depth-first traversal of the structure's graph, which identifies all of the cycles to which it belongs.
We impose a ten-atom length cutoff on this depth-first search to keep our ring search computationally reasonable.
Finally, we discard any cycle whose interior region contains one or more carbon atom which is not part of the cycle, using the ray-crossing algorithm.
We then tally the number of carbons in each of the cycles obtained by our procedure to determine the ring distributions in our generated MAC samples.

\subsection*{Distinguishing crystalline and isolated hexagons}

After having used our ring-finding method (described in the previous section) to identify the different rings in MAP-generated sAMC structures, we want to differentiate between the ``crystalline" hexagons -- which are clustered together into highly ordered, graphene-like regions -- from the ``isolated'' hexagons, which tend to be more distorted than the crystalline ones, and are usually surrounded with carbon rings of different sizes.
To do so, we begin by building a hexagon adjacency matrix $M_h$ for structure, whose elements are defined as follows:

\begin{align*}
  M_{h, ij} = \begin{cases}
    1 &\text{if } \text{ hexagons } i \text{ and } j \text{ share a vertex (i.e. a carbon atom),} \\
    0 &\text{if } \text{ otherwise.}
  \end{cases}
\end{align*}

We then use depth-first search of the structure's hexagon graph (as defined by $M_h$) to find its connected components.
We define a ``flake'' motif as a hexagon connected to six other hexagons: we consider all seven hexagons as part of the flake.
A crystalline hexagon is then defined as (i) part of a flake; or (ii) connected to a hexgon which is part of a flake.
All hexagons which do not satisfy either of these two criteria are defined as isolated hexagons.

\section{S3. Bond length and bond angle distributions}

\begin{figure}
  \includegraphics[width=1.0\textwidth]{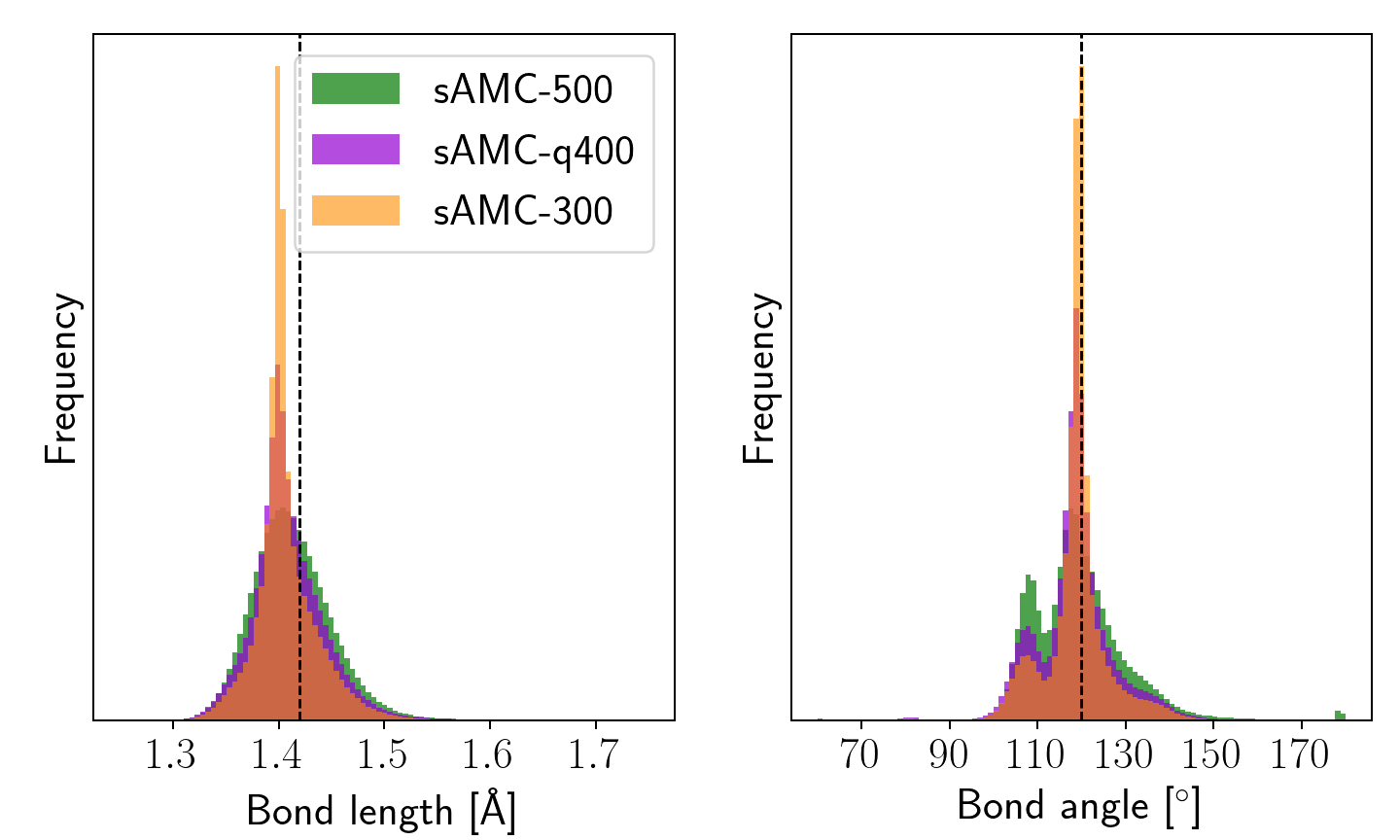}
  \caption{\textbf{Bond length and bond angle distributions of the three sAMC ensembles.} The vertical dashed lines at $r = 1.42$~\AA (\textit{left}) and $\theta = 120\degree$ (\textit{right}) correspond to the bond length and bond angle found in pristine crystalline graphene.}
  \label{fig:bond_hists}
\end{figure}

We plot the bond length and bond angle distributions of the three sAMC ensembles in Figure~\ref{fig:bond_hists}.
Both distributions are very similar for all three sets of structures: they all exhibit peaks at the same values, and, as expected, those peaks are broader in the more disordered ensembles (something that can also be observed in the experimental distributions).

Overall, the bond length (bond angle) distributions plotted in Figure~\ref{fig:bond_hists} resemble those plotted in Figure~2i (2j) of Ref.\citenum{tian2023disorder}: both simulated and experimental distributions are maximized at 1.42~\AA, (120$\degree$) corresponding to the bond length (bond angle) in crystalline graphene, and in both cases, the distributions get broader as the degree of disorder of their corresponding ensemble increases.

However, we do not notice several differences between our simulated distributions and the experimetnal ones.
Firstly, our bond angle distributions appear to be slightly right-skewed, whereas the curves in Fig.~2i of Ref.~\citenum{tian2023disorder} appear to be Gaussian.

More notably, our simulated bond angle distributions exhibits a few smaller peaks, most prominently around $109.5
\degree$ and $170\degree$, which demonstrate that a small but appreciable fraction of carbon atoms in our sAMC structures are four- and twofold coordinated, respectively.
Such peaks are absent in the experimental distribution, suggesting that carbon atoms in the experimental AMCs are universally threefold coordinated.
This might be due to biased sampling of the experimental. Indeed, inspection of some of the ADF-STEM images used by the authors to produce their distributions reveal large regions of the AMC samples that are out of focus, presumably due to small ripples in the carbon films.
Local coordination defects can induce short-wavelength ripples and buckling in graphene~\cite{ma2009stone}.
It is therefore possible that these unresolved sections of the experimental AMC samples contain carbon atoms that are not threefold coordinated, and, if accounted for, would reconcile the differences between the experimental and simulated bond angle distributions.

Finally, we would like to add that coordination defects and strained bonding configurations have roughly the same effects on the outcome of a tight-binding electronic structure calculation with orthogonal atomic orbitals: both types of structural deformations induce near-$\epsilon_F$ defect-localised states~\cite{fedders1992defects,chen2021stone}.
So, while it \emph{could} be possible that the experimental samples described in Ref.~\citenum{tian2023disorder} are completely devoid of coordination defects, their broadened bond angle and bond length distributions, taken with ADF-STEM images presented therein, indicate the presence of many strained bonds.
Such coordination-preserving structural defects (\textit{\`{a} la} Stone-Wales) will yield tight-binding eigenstates similar to those we obtained from our simulated structures.
We therefore do not expect our method to yield appreciably different results on structures with bond angle and bond length distributions identical to those reported in Ref.~\citenum{tian2023disorder}.

\section{S4. Electronic structure: tight-binding hamiltonian}

Starting from a AMC fragment with carbon atoms at positions $\{\mathbf{R}_i\}_{i=1,\,\ldots ,\, N}$, we dress each carbon with a $2p_z$ atomic orbital (AO) $|\varphi_i\rangle$.
We then construct its tight-binding hamiltonian $\mathcal{H}$ whose elements are defined using the Linderberg approximation~\cite{linderberg1967consistency}:

\begin{equation*}
    \langle\varphi_i|\mathcal{H}|\varphi_j\rangle = t_{ij} = 
    \begin{cases}
        \beta_0\,e^{-\mu_b(R_{ij} - R_0)}\,(1 + k_b(R_{ij} - R_0)) &\iff 0 < R_{ij} = |\mathbf{R}_i - \mathbf{R}_j| \le R_{\text{cutoff}}\,, \\ 
        0 & \iff R_{ij} > R_{\text{cutoff}}\,,
    \end{cases}
\end{equation*}

where $\beta_0$, $\mu_b$, and $k_b$ are semi-empirical parameters describing $sp^2$ carbons, and $R_0$ is the equilibrium bond length between two carbon $\pi$-conjugated carbon atoms.
The numerical values of these parameters can be found in Table~S1.
We set the nearest-neighbour cutoff distance  $R_{\text{cutoff}} = 1.8\,$\AA.
The parametrisation of our tight-binding hamiltonian was adopted from previous publications~\cite{sterpone2008molecular,jailaubekov2013hot}. 

\begin{table}[h]
  \centering
	\caption{\textbf{Tight-binding hamiltonian parameters}}
	\label{tab:tb_params} 
  \begin{tabular}{cccc}
  \hline
  $\beta_0$ {[}eV{]} & $\mu_b$ {[}\AA$^{-1}${]} & $R_0$ {[}\AA{]} & $k_b$ {[}\AA$^{-1}${]} \\ \hline
  -2.438             & 2.035                                   & 1.397                          & 0.405                                
  \end{tabular}
\end{table}

\section{S5. Partial diagonalisation of tight-binding hamiltonians}

Every AMC structure studied in this work contains  $N\approx 60000$ carbon atoms.
The hamiltonians we are working with are therefore massive matrices, whose diagonalisation using standard numerical techniques is completely intractable.
However, each carbon atom typically has between two and four neighbours within $R_{\text{cutoff}}$, meaning that $\mathcal{H}$ contains mostly null entries.
We therefore represent each structure's hamiltonian as a sparse matrix to reduce the memory and time costs of its diagonalisation.

The full eigenspectrum of very large sparse matrices cannot be resolved with a single diagonalisation routine.
Instead, specialised algorithms exist to obtain relatively small subsets of its eigenvector-eigenvalue pairs that occupy the extremal portions of its spectrum.
The same algorithms can also find the eigenpairs that lie closest to (either above or below) some user-defined constant $\lambda$, using the shift-and-invert technique~\cite{ericsson1980spectral}.

We use the Lanczos algorithm, as implemented in \textsc{ARPACK} library, to compute the different subsets of tight-binding eigenstates expected to be thermally accessible to charge hopping in the three gating regimes considered in this work.
For the high postitve (negative) gate voltage regime, we use the unshifted alogrithm to obtain the 100 highest-energy (lowest-energy) eigenstates and keep only the eigenstates whose energies lie within $4 k_{\text{B}}T$ of the maximum (minimum) energy, at $T = 300$~K.


For the ungated regime, we only consider the unoccupied eigenstates within $4 k_{\text{B}} T$ of each sample's Fermi energy.
To do so, we use the shift-inverted Lanczos algorithm with $\lambda = \epsilon_{\text{HOMO}}$, the energy of the highest occupied molecular orbital (HOMO), which we must estimate for each fragment.
First, we use Gershgorin's circle theorem~\cite{higham2008functions} to estimate the spectral range of $\mathcal{H}$:

\begin{equation*}
  \forall n\in\{1,\ldots ,N\},\exists i\in\{1,\ldots ,N\},\, |\epsilon_n - \mathcal{H}_{ii}| \le \sum_{j = 1}^N |t_{ij}| = r_i\,,
\end{equation*}
where $\epsilon_n$ denotes the n$^{th}$ eigenvalue of $\mathcal{H}$, and $t_{ij}$ are its off-diagonal elements.
In words, the theorem states that any eigenvalue $\epsilon_n$ of $\mathcal{H}$ lies within the range $r_i$ of some diagonal element $\mathcal{H}_{ii}$.

Looping over all $N$ rows (as indexed by $i$) in $\mathcal{H}$, we can obtain an interval $[E_{min}; E_{max}]$ which must contain all of its eigenvalues, where $E_{min} = \min_{i} \{H_{ii} - r_i\}$ and $E_{max}=\max_{i} \{H_{ii} + r_i\}$.
We use this interval to produce an initial guess of the HOMO energy $\epsilon^{(0)} = (E_{min} + E_{max})/2$.
Knowing that $\mathcal{H}_{ii} = 0$ for all $i$ implies $E_{min} = -E_{max}$, which simplifies our initial guess $\varepsilon^{(0)} = 0$~eV for all AMC structures.

Next, we perform a cheap LDLT factorisation:
\[\mathcal{H} - \varepsilon^{(0)} = LDL^{\text{T}}\,\]
and count the number of negative entries of the diagonal matrix $D$ to determine the number $n_<^{(0)}$ of eigenvalues $\epsilon_i < \epsilon^{(0)}$.
Our model of AMC's electronic structure only accounts for a single electron per carbon atom and assumes spin degeneracy, so the HOMO energy will have exactly $N/2$ energies inferior or equal to it.
We then define:
\[\delta^{(0)} = n_<^{(0)} - N/2\,,\]
and obtain a better estimate of $\epsilon_{\text{HOMO}}$ by running ARPACK's Lanczos routine in eigenvalue-only mode (saves time and memory) with a low accuracy threshold ($\varepsilon_{tol} = 10^{-5}$), and requiring $\delta^{(0)}$ eigenvalues.

We then loop over all approximate eigenvalues $\{\epsilon_i^{(1)}\}_{i=1,\,\ldots ,\,\delta^{(0)}}$ we obtain and estimate the number $n^{(1)}_{<,i}$ of true eigenvalues smaller than $\epsilon_i^{(1)}$, using the same LDL factorisation method as above.
If one of our eigenvalue estimates $\epsilon_{i}^{(1)}$ satisfies $n_{<,i}^{(1)} = N/2$, we take it as our estimate $\epsilon_{\text{HOMO}}$.
Otherwise, we re-run the approximate diagonalisation of $\mathcal{H}$, until at some iteration $n$, we find a satisfactory eigenvalue estimate which satisfies $n_<^{(n)} = N/2$.
We stress that the crude eigenvalue estimates we obtain at this point of our algorithm only serve to aim ARPACK's diagonalisation routine to the correct region of each structure's spectrum: we do not interpret them as useful physical quantities in our model.

Having obtained a rough estimate of $\epsilon_{\text{HOMO}}$, we compute the number of $\Delta$ of energy eigenstates lie between $\epsilon_{\text{HOMO}}$ and $\epsilon_{\text{HOMO}} + 4\,k_{\text{B}} T$, with $T = 300\,$K, using the LDLT factorisation of $\mathcal{H} - (\epsilon_{\text{HOMO}} + 4k_{\text{B}} T)$.
We then run ARPACK's Lanczos routine in eigenvalue-and-eigenvector mode with a tight convergence criterion ($\varepsilon_{tol} = 10^{-8}$), $\lambda=\epsilon_{\text{HOMO}}-\delta\epsilon$, and ask for $\Delta+1$ eigenpairs above $\lambda$.
This finally yields between one and two hundred molecular orbitals $\{|\psi_{\text{HOMO}n}\rangle\}_{n = 0,\,\ldots ,\,\Delta}$ and their associated energies $\{\epsilon_{\text{HOMO} + n}\}_{n=0,\,\ldots ,\,\Delta}$.

Finally we compute the Fermi energy as follows: $\epsilon_{\text{F}} = (\epsilon_{\text{HOMO}} + \epsilon_{\text{LUMO}})/2$, where $\epsilon_{\text{LUMO}} = \epsilon_{\text{HOMO}+1}$ is the second lowest energy (after $\epsilon_{\text{HOMO}}$) obtained from this procedure.
We do not use $|\psi_{\text{HOMO}}\rangle$ in our hopping calculations.

\section{S6. Comparison of our diagonalisation method with exact benchmark}

We test the accuracy of our approximate tight-binding diagonalisation scheme on a truncated sAMC-500 structure containing 5862 atoms, which is small enough to obtain its full eigenspectrum without resorting to our Lanczos-based method.
We construct its tight-binding Hamiltonian using the method outlined in Section S3 of the manuscript's Supporting Information, and diagonalise it exactly using the LAPACK linear algebra package.
We further compute its approximate near-$\epsilon_F$, lowest-, and highest-energy molecular orbitals using our protocol described in Section S4 of the Supporting Information.

\begin{figure}
  \includegraphics[width=1.0\textwidth]{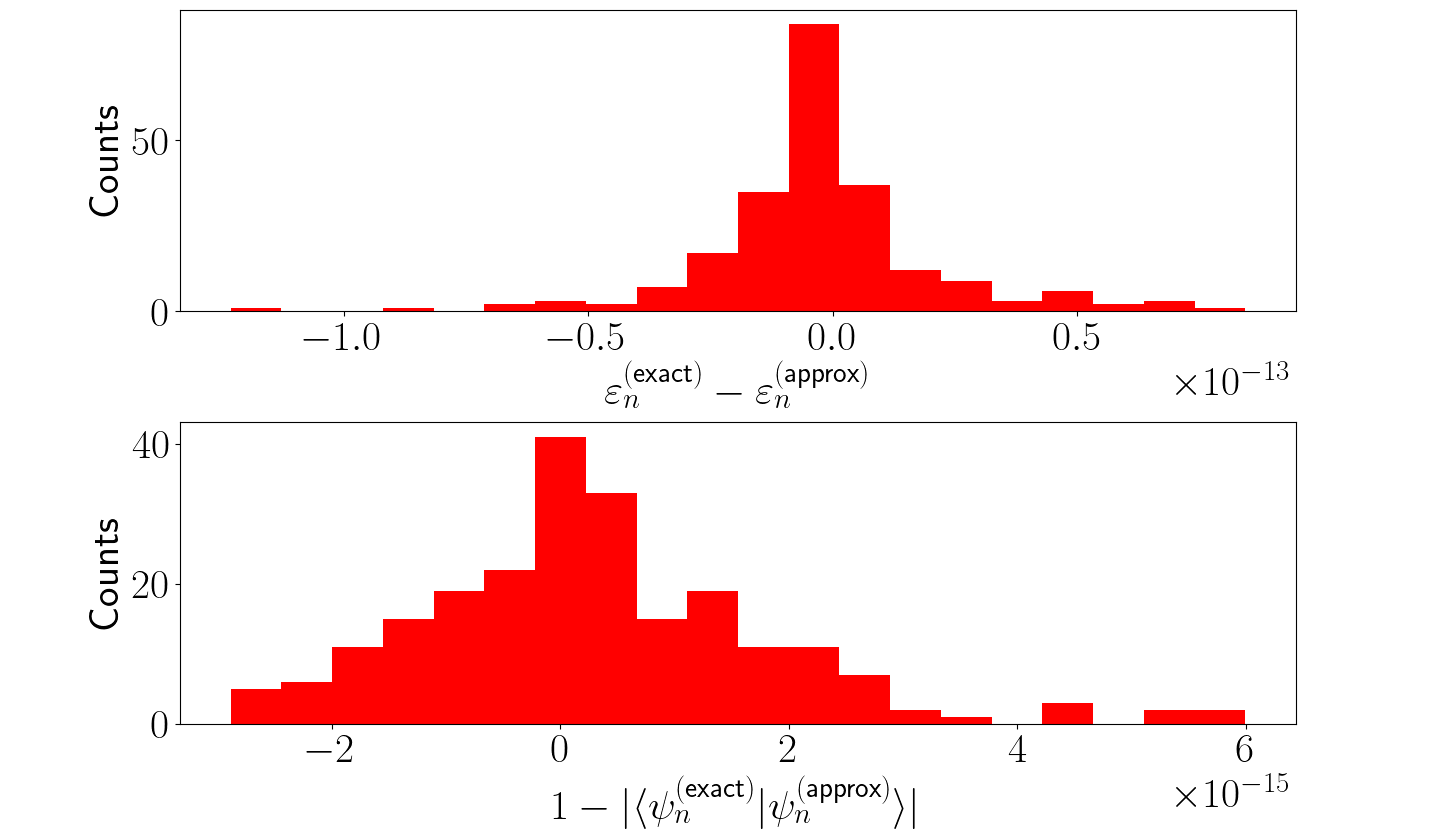}
  \caption{\textbf{Comparison of our Lanczos-based tight-binding diagonalisation scheme with exact benchmark on small structure.} \textit{Top:} Distribution of differences between approximate tight-binding eigenvalues and those obtained using a numerically exact diagonalisation scheme. \textit{Bottom:} Distribution of devation from unity of the inner products between the approximate tight-binding eigenstates from our method and their exact analogs.}
  \label{fig:compare_w_exact}
\end{figure}

We compare the results of the two diagonalisation method by computing the energy differences between the exact energies and their approximate counterparts, as well as the deviations from unity of the inner products\footnote{More precisely we examine $1-|\langle\psi_n^{(\text{exact})}|\psi_n^{(\text{approx})}\rangle|$, because the eigenvectors obtained by the two diagonalisation schemes sometimes differ by an overall sign of -1.} between the two sets of eigenvectors in Figure~\ref{fig:compare_w_exact}.
This comparison is plotted in Figure 1, which shows that both sets energies and molecular orbitals are equivalent, up to machine accuracy (i.e., $\sim10^{-13}$).

Comparing our percolation-based results with a higher level of theory is less straightforward. While there exists several more exact approaches to simulate charge transport in materials (e.g., non-equilibrium Green's functions), these methods describe coherent charge transport, which is a different physical mechanism. We therefore do not expect such methods to yield predictions which are directly comparable with those of our percolation-based approach.

Other approximate schemes have been used to simulate incoherent, variable range hopping, and have had reasonable agreement with experiment.
There does not appear to be a widely accepted benchmark method for these types of simulations.

\section{S7. Constructing the variable range hopping network sites} 

\subsection*{Procedure for localizing distributed molecular orbitals}
Here we describe the procedure for localizing the molecular orbitals (MOs) which were found to exhibit a torn morphology with multiple disjoint high-density pockets (see left panel of Fig.~S1). 
For reasons discussed in the main text, we extract hopping sites from a tight-binding molecular orbital (MO) $|\psi_n\rangle$ by partitioning the atoms in the sAMC sample using a $k$-means clustering approach. For each MO $|\psi_n\rangle$, we determine the number of clusters by estimating the number of high-electronic-density pockets it admits (more details in the following section), and proceed to group the atoms in the AMC sample using Lloyd's algorithm~\cite{lloyd1982least,scikit-learn}, weighed by its electronic density squared: $|\langle\varphi_j|\psi_n\rangle|^4$.
Each cluster defines a new electronic state which is much more compactly localised than the original MO (see middle panel of Figure~S1).
We take these states to be the hopping sites, and define their respective energies as the energies of their mother MO.
For a given site $|s_i\rangle$, we define its position as: 

\begin{equation}
  \tilde{\mathbf{R}}_i = \frac{\sum_j \mathbf{r}_j |\langle\varphi_j|s_i\rangle|^4 }{\sum_j |\langle\varphi_j|s_i\rangle|^4}\,,
  \label{eq:site_pos}
\end{equation}

where $\mathbf{r}_j$ denotes the position of carbon atom $j$.
We use the square of the site's electronic density $|\langle\varphi_j|s_i\rangle|^4$ for the clustering of the sites and our evaluation of $\mathbf{\tilde{R}}_i$ to ensure that the hopping sites produced by this protocal are centered on or near a pocket of high electronic density.

\begin{figure}[h]
  \centering
  \includegraphics[width=1.15\textwidth]{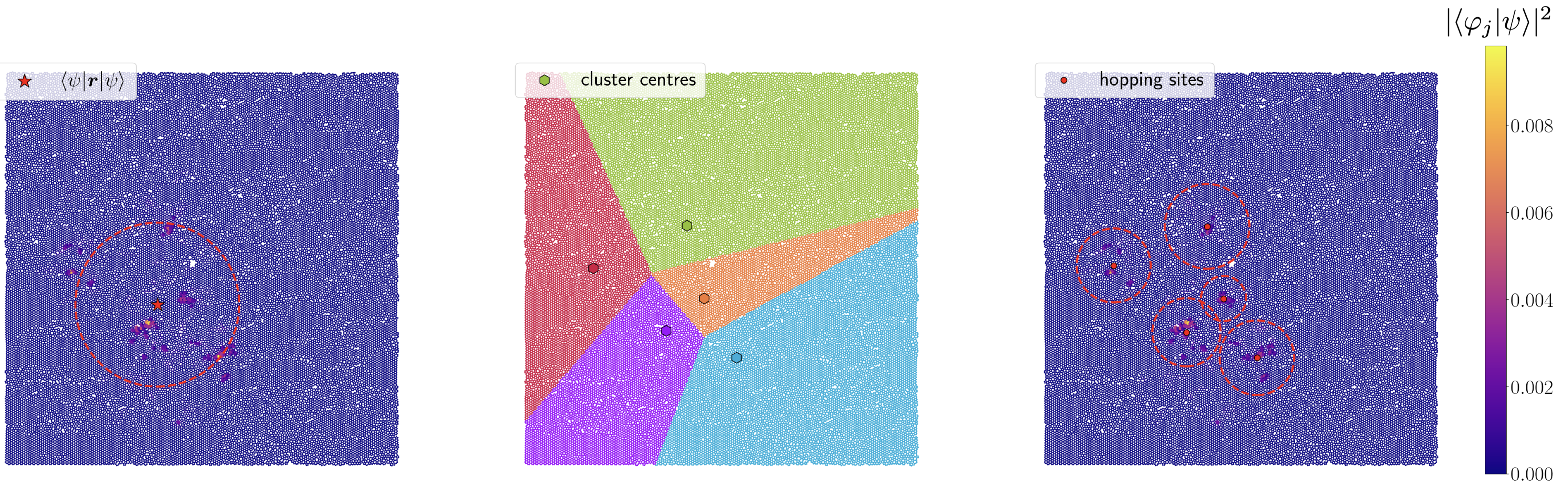}
  \caption{\textbf{Procedure for obtaining the hopping sites}. \textit{Left:} Example of a molecular orbital (MO) $|\psi\rangle$ obtained by diagonalising the tight-binding hamiltonian $\mathcal{H}$. Atomic orbitals $|\varphi_j\rangle$ for whom $|\langle\varphi_j|\psi\rangle|^2\ge0.001$ are enlarged. The MO's center of mass $\langle\textbf{r}\rangle$ is displayed as a star, and its radius of gyration $\sqrt{\langle\textbf{r}^2\rangle - \langle\textbf{r}\rangle^2}$ is shown in red.
  \textit{Middle:} Partition of the atoms in the AMC sample using our MO-weighted k-means clustering. The hexagons mark each cluster's center of mass.
  \textit{Right:} Positions (dots) and radii (dotted circles) of the final hopping sites obtained from $|\psi\rangle$. The green cluster in \textbf{(B)} is devoid of any high electronic density pocket, so it does not yield a hopping site.
  The color bar applies to the left and right Figures and measures the electronic density. The dimensions of the samples are 40~nm$\times$40~nm.}
  \label{fig:hopping_sites}
\end{figure}

We allow each structure to have sites with variable radii, which we define as:

\begin{equation}
  a_i = \sqrt{\langle s_i|\,\mathbf{r}^2\,|s_i\rangle - \langle s_i|\,\mathbf{r}\,|s_i\rangle}\,,
\end{equation}

with $\mathbf{r}$ denoting the position operator. 
The positions and radii of the sites obtained by this procedure are shown in the right panel of Figure~S1.

\subsection*{Identifying the number of clusters}

We define our hopping sites by running $k$-means clustering on the MOs we obtain from our partial diagonalisation routine.
The number of clusters into which the algorithm partitions a given MO must be specified before running it.
We define the number of clusters for each MO by using a crude estimate of the number of local maxima it exhibits.

\begin{figure}[h]
  \centering
  \includegraphics[width=1.1\textwidth]{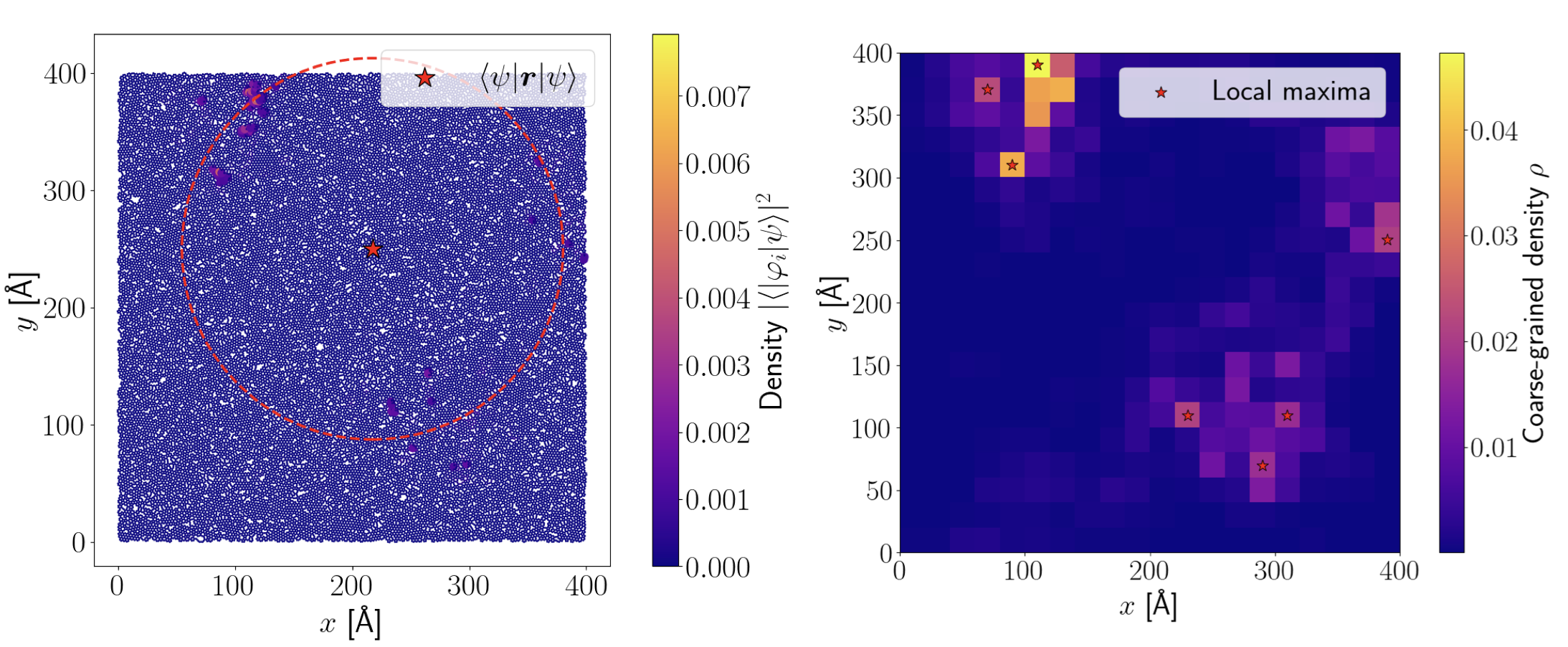}
  \caption{\textbf{Procedure for determining the number of hopping sites defined by a tight-binding molecular orbital}. \textit{Left:} Example of a molecular orbital (MO) $|\psi\rangle$ obtained by diagonalising the tight-binding hamiltonian $\mathcal{H}$. The red star represents the MO's average position $\langle\textbf{r}\rangle = \langle\psi|\textbf{r}|\psi\rangle$ and the red dotted circle centered on $\langle\textbf{r}\rangle$ delimits the radius of gyration of $|\psi\rangle$ defined as: $R_{g} =\sqrt{\langle\psi|\textbf{r}^2|\psi\rangle - |\langle\textbf{r}\rangle|^2}$.
  \textit{Right:} Coarse-grained representation of $|\psi\rangle$ on a grid of $20\times20$ bins $\mathcal{B}_{pq}$. A coarser grid is used here for the sake of illustration. The bins corresponding to approximate local maxima are marked with a red star. In this example, the algorithm has identified 7 local maxima, so our $k$-means clustering protocol will parition $|\psi\rangle$ into at most 7 sites.} 
  \label{fig:gridify_mos}
\end{figure}  

For a given AMC fragment (for example see left panel in Fig.~S2), our diagonalisation routine yields each eigenstate $|\psi_n\rangle$ as vectors in the basis of the carbon atoms' $2p_z$ orbitals $\{|\varphi_j\rangle\}_{j=1,\,\ldots ,\,N}$ : 
\[|\psi_n\rangle = \sum_{j=1}^{N} |\varphi_j\rangle\langle\varphi_j|\psi_n\rangle\,.\]
We project this representation back into real space by dividing the amorphous graphene into a grid of 100$\times$100 bins $\{\mathcal{B}_{pq}\}_{p,q=1,\ldots,100}$, and assigning to each bin $\mathcal{B}_{pq}$ the coarse-grained electronic density value: 
\[\rho(p,q) = \sum_{\mathbf{R}_j\in\mathcal{B}_{pq}}|\langle\varphi_j|\psi_n\rangle|^2.\]
We then identify the centers of the bins that locally maximise $\rho$ (plotted in the right panel of Fig.~S2) and obtain their real-space coordinates.
To avoid over-partitioning our MOs, we ignore all bins $\mathcal{B}_{pq}$ for which $\rho(p,q) < 0.3\cdot \max_{p,q} \{\rho(p,q)\}$.
We also impose a minimum distance of 20\,\AA between coarse-grained maxima we obtain using this procedure to avoid assigning multiple local maxima to a single high-density pocket.

\section{S8. Numerical implementation of percolation theory}

We describe hopping using a symmetric Miller-Abrahams rate expression~\cite{miller1960impurity}, adapted to accommodate sites with different radii:

\begin{equation}
  \omega_{ij} = \omega_0 \exp\left\{ -f(|\mathbf{R}_i - \mathbf{R}_j|, a_i, a_j)  - \frac{|\varepsilon_i - \mu| + |\varepsilon_j - \mu| + |\varepsilon_i - \varepsilon_j|}{2k_{\text{B}}T} \right\} = \omega_{0} e^{-\xi_{ij}}\,,
  \label{eq:MA_rates}
\end{equation}

where $\omega_0 = 10^{15}$~s$^{-1}$ is the attempt-to-escape frequency, $i$ and $j$ label the hopping sites, $\mathbf{R}_i$ and $\mathbf{R}_j$ denote their positions, $a_i$ and $a_j$ denote their radii, and $f(|\mu_1 - \mu_2|, a_1, a_2)$ corresponds to the overlap between two Gaussians with means $\mu_i$, and standard deviations $a_i$:

\begin{equation}
  f(|\mu_1 - \mu_2|, \sigma_1, \sigma_2) = \frac{\sigma_1\,\sigma_2}{2\pi\left(\sigma_1^2 + \sigma_2^2\right)}\,\exp\left[ -\frac{|\mu_1 - \mu_2|^2}{2\left(\sigma_1^2 + \sigma_2^2\right)} \right]\,.
  \label{eq:overlap}
\end{equation}

To calculate the conductance $G$ of an AMC ensemble, we must determine the percolation threshold $\xi_c$ for each structure in the ensemble.
For each structure, we iterate over its inter-site distances sorted in ascending order $\xi^{(1)}\le\xi^{(2)}\le\ldots$, connecting all pairs of sites of $(i,j)$ such that $\xi_{ij} \le \xi^{(k)}$, on a given iteration step $k$.
We halt the process once we have constructed a network of sites that connects a site strongly coupled to the left electrode, to a site strongly coupled to the right electrode. 
We then record the distance $\xi_c$ at which a percolating cluster of sites emerges for that specific structure.
We use the distribution of percolation thresholds we obtained for each ensemble to compute their respective conductances.

We estimate the uncertainty $\sigma_{G}$ associated with the conductance by ablation.
For each ensemble, we re-compute the conductance by omitting a single sAMC sample from the ensemble, and we repeat this operation for each structure in the ensemble.
We then define the uncertainty in our estimate of the ensemble's conductance as follows:

\begin{equation*}
    \sigma_G = \max_{n} |G - G^{(n)}|\,,
\end{equation*}
where $G$ is the conductance we obtained by including all structures in the ensemble in our calculation, and $G^{(n)}$ is the conductance obtained when omitting only the n$^{th}$ structure from the percolation calculation.

\section{S9. Crystallinity distributions of tight-binding eigenstates in different regions of the energy spectrum}

\begin{figure}
  \includegraphics[width=1.0\textwidth]{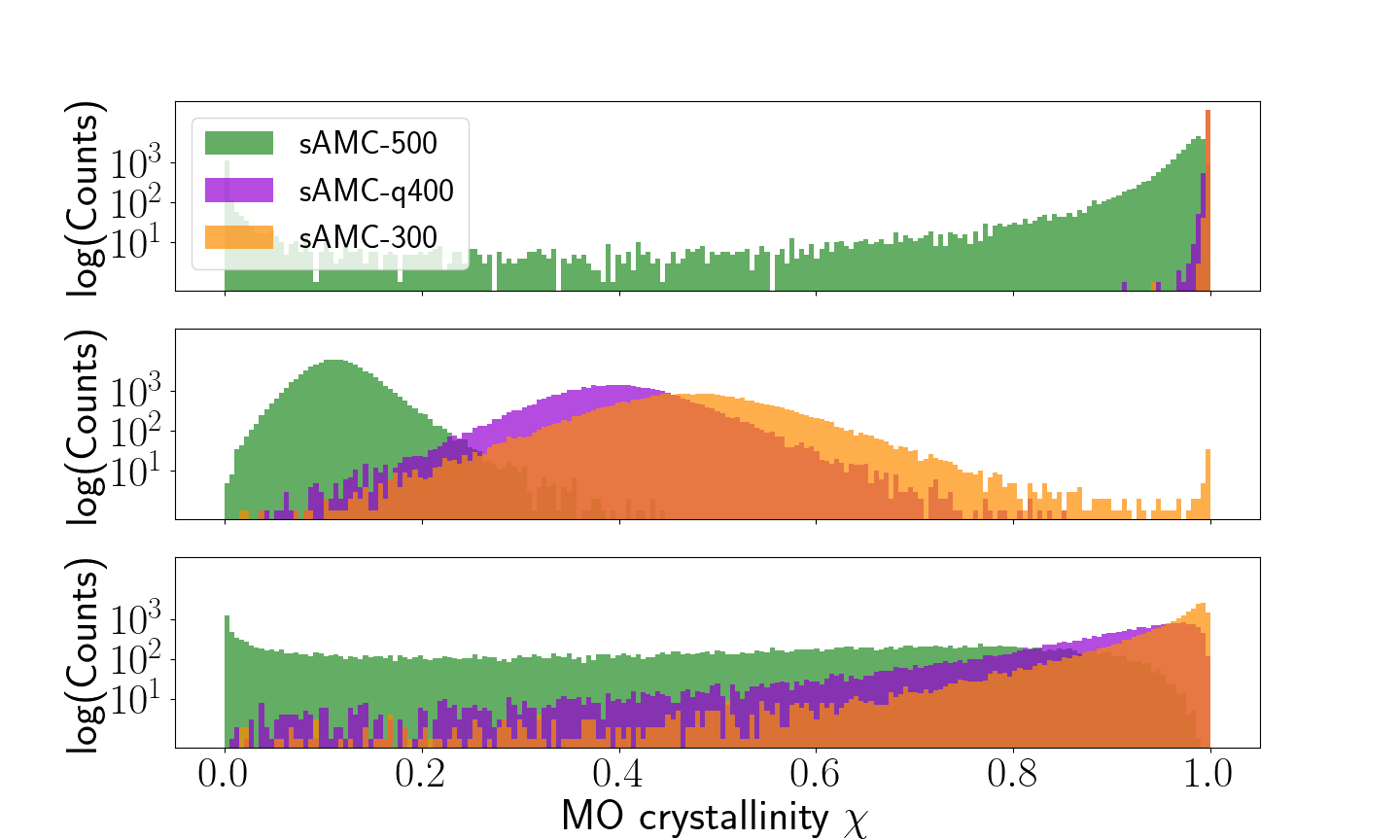}
  \caption{\textbf{Crystallinity distributions of the tight-binding molecular orbitals (MOs) in different regions of the energy spectrum}. \textit{Top:} MOs with energies within $4k_{\text{B}}T$ of each fragment's highest energy state. \textit{Middle:} Lowest-lying virtual MOs within $4k_{\text{B}}T$ of each fragment's Fermi energy. \textit{Bottom:} MOs with energy within 4$k_{\text{B}}T$ of each fragment's lowest energy state. Note that all ordinate axes are logarithmically scaled.}
  \label{fig:MO_cryst}
\end{figure}

The crystallinity of the conducting sites is related to the crystallinity of the molecular orbitals (MOs) from which they are obtained.
As can be seen in Figure~\ref{fig:MO_cryst}, the crystallinity distributions of the extremal MOs of sAMC-500 have much broader support on $[0,1]$ (green on Figure 3) than those of ensembles sAMC-q400 and sAMC-300 (respectively purple and orange on Figure 3), which exhibit highly crystalline MOs in the extreme regions of their spectra.
The MO crystallinity distributions plotted in Figure~\ref{fig:MO_cryst} closely match the distributions of the hopping sites plotted in Figures~3b-d in the main text, thus offering a first layer of explanation of the divergences in the crystallinity distributions of the sites.

Although it is difficult to rigorously establish a universal link between an MO's crystallinity and its energy, our observations are consistent with the well-established theory of the electronic structure of disordered semiconductors.
Structural disorder is known to induce electronic localisation: disordered systems have been shown to support both delocalised and localised states, separated by a mobility edge~\cite{anderson1958absence,economou1970localization}.
Delocalised states occupy the extreme regions of the density of states, and the localized states lie in its centre, near the Fermi energy~\cite{mott2012electronic}.
Consider first the weakly disordered limit of a crystalline semiconductor, with a few isolated defects (e.g. doping agents, vacancies, lattice defects).
Such systems are known to admit defect-localised electronic states near and within their bandgap, while the electronic states deeper in their conduction and valence bands are unperturbed by the defects~\cite{shklovskii1984electronic}, and therefore remain highly crystalline.
In the opposite limit, we have Zachariasen continuous random networks (Z-CRNs), which exhibit a total lack of long-range order, but are statistically homogeneous and isotropic.
A delocalised electronic state supported by a Z-CRN is likely to sample a diverse set of local bonding configurations due to the randomness of the bonding network~\cite{dong1998atomistic,ludlam2005universal}.
The AMC structures considered in our manuscript lie somewhere between these two limiting cases on the disorder spectrum.

On one hand, we have the more ordered sAMC-300 and sAMC-q400 ensembles, which both exhibit sufficiently extended crystalline domains that they could almost be regarded as heavily defected graphene. Consequently, the states deep within their respective manifolds of virtual and occupied MOs (i.e., their conduction and valence ``bands") retain a highly crystalline character, while their near-$\epsilon_F$ MOs are concentrated near ``defects": the disordered regions.

The case of sAMC-500 is a bit more subtle, its crystalline regions are much smaller than those found in sAMC-q400 and sAMC-300. Its structure is closer to a Z-CRN.
The tight-binding eigenstates of its structures can therefore be expected to homogeneously sample their supporting structures, thus explaining their roughly uniform crystallinity distribution. 
In this light, the penchant of sAMC-500's conducting sites for amorphous for disordered regions in the ungated regime (Fig. 3d) is a nontrivial result.

\section{S10. Code availability}
The code used for to instantiate and partially diagonalise the tight-binding hamiltonians can be found in the repository listed as reference~\citenum{Gastellu_ArpackMAC}.

Our numerical percolation code may be found in the repository listed as reference~\citenum{Gastellu_PercolateMAC}.

The code used for the structural analysis of MAP-generated AMC samples can be found in the repository listed as reference~\citenum{Gastellu_DisorderAnalysisMAC}. The XYZ coordinate files describing the AMC structures used in our study can be also be found in this repository. These XYZ files can be used as input for the tight-binding and percolation codes mentioned above.

The code used for the training and deployment of MAP used to generate our three AMC ensembles can be found in the repository listed as reference~\citenum{Gastellu_sAMC_generation}.

\bibliography{refs}